\documentclass[aps,prd,twocolumn,preprintnumbers,amsmath,amssymb,
superscriptaddress,showpacs,floatfix]{revtex4}

\usepackage{graphicx}
\usepackage{dcolumn}
\usepackage{bm}
\usepackage{epsfig}
\usepackage{graphics}
\usepackage{latexsym}

\def\beq{\begin{equation}}
\def\eeq{\end{equation}}
\def\bea{\begin{eqnarray}}
\def\eea{\end{eqnarray}}
\newcommand{\beqs}{\begin{subequations}}
\newcommand{\eeqs}{\end{subequations}}

\newcommand{\cref}[1]{Ref.~\cite{#1}}

\newcommand{\vev}[1]{\left<#1\right>}

\newcommand{\hh}{{\ensuremath{I{\kern-2.6pt h}}}}
\newcommand{\bhh}{{\ensuremath{\bar{I{\kern-2.6pt h}}}}}

\newcommand{\ka}{\kappa}
\newcommand{\la}{\lambda}
\newcommand{\Pb}{\bar{\Phi}}
\newcommand{\Hb}{\bar{H}}

\begin{document}

\preprint{UT-STPD-15/02}

\title{Gravitational Waves from Double Hybrid Inflation}

\author{G. Lazarides}
\email{lazaride@eng.auth.gr} \affiliation{School of Electrical and
Computer Engineering, Faculty of Engineering, Aristotle University
of Thessaloniki, Thessaloniki 54124, Greece}
\author{C. Panagiotakopoulos}
\email{costapan@eng.auth.gr}
\affiliation{School of Rural and Surveying Engineering, Faculty of 
Engineering, Aristotle University of Thessaloniki, Thessaloniki 
54124, Greece}

\date{\today}

\begin{abstract}
We present a two stage hybrid inflationary scenario in 
non-minimal supergravity which can 
predict values of the tensor-to-scalar ratio of the order of 
${\rm few}\times 10^{-2}$. For the parameters considered, the 
underlying supersymmetric particle physics model possesses two 
inflationary paths, the trivial and the semi-shifted one. The 
trivial path is stabilized by supergravity 
corrections and supports a first stage of inflation with a 
limited number of e-foldings. The tensor-to-scalar ratio can 
become appreciable while the value of the scalar spectral 
index remains acceptable as a result of the competition 
between the relatively mild supergravity corrections and the 
strong radiative corrections to the inflationary potential. 
The additional number of e-foldings required 
for solving the puzzles of hot big bang cosmology are 
generated by a second stage of inflation taking place 
along the semi-shifted path. This is possible only because
the semi-shifted path is almost perpendicular to the trivial one 
and, thus, not affected by the strong radiative corrections 
along the trivial path and also because the supergravity effects 
remain mild. The requirement that the running of the scalar 
spectral index remains acceptable limits the possible values of 
the tensor-to-scalar ratio not to exceed about $5\times 10^{-2}$. 
Our model predicts the formation of an unstable string-monopole 
network, which may lead to detectable gravity wave signatures in 
future space-based laser interferometer observations.
\end{abstract}

\maketitle

\section{Introduction}

Inflation (for a review see e.g. Ref.~\cite{lectures}) 
is by now considered to be an integral part of standard 
cosmology thanks to a plethora of precise observations on the 
cosmic microwave 
background radiation (CMBR) and the large-scale structure in the 
universe. Therefore, it is very important to construct realistic 
inflationary models based on particle theory and 
consistent with all the available cosmological and phenomenological 
requirements. 
Undoubtedly, hybrid inflation \cite{linde} is one of the most 
promising inflationary scenarios. It is \cite{cop,dss} 
naturally realized in the context of supersymmetric (SUSY) grand 
unified theory (GUT) models based on gauge groups with rank 
greater than or equal to five. 

In standard SUSY hybrid inflation, however,
the GUT gauge symmetry is spontaneously broken only at the end 
of inflation and, thus, if magnetic monopoles are predicted by 
this symmetry breaking, they are copiously produced 
\cite{smooth}, leading to a cosmological catastrophe. This 
disaster is avoided in the smooth \cite{smooth} 
or shifted \cite{shift} variants of SUSY hybrid inflation, 
where the GUT gauge symmetry is broken already during 
inflation. These variants were based on non-renormalizable 
superpotential terms. It was, though, subsequently shown that 
a new smooth \cite{nsmooth} and a new shifted \cite{nshift} 
hybrid inflation scenario can be constructed with only 
renormalizable superpotential terms within an extended 
Pati-Salam (PS) SUSY GUT model, which was initially 
introduced \cite{quasi} for solving a very different 
problem. Namely, the simplest SUSY PS model predicts (see 
Ref.~\cite{hw}) exact Yukawa unification \cite{als} and, if 
it is supplemented with universal boundary conditions, yields
unacceptable $b$-quark mass values. This problem is 
solved in the extended model, where Yukawa unification is 
naturally and moderately violated.

After the first accurate measurement \cite{wmap07} of 
the scalar spectral index $n_{\rm s}$, however, it has been 
realized that there is a tension between all 
these well-motivated and natural inflationary scenarios and 
the measured value of this index. Indeed, within the 
standard power-law cosmological model with cold dark matter 
and a cosmological constant, the data imply 
that $n_{\rm s}$ is clearly lower than unity -- for 
the latest results on $n_{\rm s}$ see Ref.~\cite{planck15}. 
Inflationary scenarios, on the other hand, such as the ones 
mentioned above, within supergravity (SUGRA) with minimal 
K\"{a}hler potential, yield \cite{senoguz} $n_{\rm s}$'s 
which are very close to unity or even exceed it.

One idea \cite{mhin} for reducing the predicted spectral index 
is based on the observation that 
$n_{\rm s}$ generally decreases with the number of e-foldings 
suffered by our present horizon scale during inflation. So, 
reducing this number of e-foldings, we can achieve values of 
$n_{\rm s}$ compatible with the recent data without having to
abandon the use of a minimal K\"{a}hler potential. The
additional number of e-foldings required for solving the 
horizon and flatness problems of standard hot big bang 
cosmology can be provided by a subsequent second stage of 
inflation. 

It is interesting to note that the extended SUSY PS 
model of Ref.~\cite{quasi}, which can lead to new smooth 
\cite{nsmooth} or new shifted \cite{nshift} hybrid inflation, 
can also provide us with a double inflation scenario called 
standard-smooth hybrid inflation \cite{stsmhi} which solves 
the above mentioned spectral index problem along the lines just 
discussed. The cosmological scales exit the horizon during the 
main stage of inflation, which is of the standard hybrid type 
and occurs as the system slowly rolls down a trivial classically 
flat direction on which the PS gauge group is unbroken. This 
direction is subsequently destabilized giving its place to a 
classically non-flat valley of minima along which new smooth 
hybrid inflation takes place with the PS GUT gauge group being
broken. Consequently, magnetic monopoles are produced only at 
the end of the first stage of inflation, but they are adequately 
diluted by the second stage, which also provides the extra 
e-foldings needed for solving the puzzles of hot big bang 
cosmology.
   
After the recent results of BICEP2 \cite{bicep2} on the B-mode 
in the polarization of the CMBR at degree angular scales, it 
seems possible that the inflationary scenarios will have to face 
a new challenge. Namely, they should be able to accommodate 
appreciable values of the tensor-to-scalar ratio $r$, since a 
B-mode of primordial origin could be due to the production of 
gravitational waves during inflation. We should, however, 
consider this possibility with reservation since some serious 
criticism \cite{criticism} to the original BICEP2 analysis 
has already appeared claiming that the foreground from Galactic 
polarized-dust emission has been underestimated. On the other 
hand, after the recently released Planck HFI 353 GHz dust 
polarization data \cite{planck}, the first attempts to make a 
joint analysis of the Planck and BICEP2 data have been 
presented \cite{joint1,joint2}. They showed that, although $r$ 
is smaller than initially claimed, significant values of $r$ -- 
of order 0.01 -- cannot be excluded. The most recent joint 
analysis \cite{joint2} yields an upper limit on $r$ of about 
0.12 at $95\%$ confidence level. Unfortunately, all 
the above mentioned variants of SUSY hybrid inflation predict 
negligible values of $r$. So, it is certainly worth 
investigating whether realistic SUSY hybrid inflation models 
accommodating appreciable values of $r$ can be constructed.

In Ref.~\cite{seto}, a double inflation scenario has been 
proposed which is compatible with the BICEP2 data 
\cite{bicep2}. The first stage of inflation is of the SUSY 
hybrid type, while the second stage is left unspecified, 
which makes the scenario incomplete. The inflationary 
potential is supplemented with a mass-squared term for the 
inflaton attributed to SUGRA corrections and with a 
logarithmic term representing very strong radiative 
corrections due to the SUSY breaking during inflation. It 
is the competition between these two contributions which 
allows appreciable values of $r$ while $n_{\rm s}$ remains 
acceptable. The assumption, however, 
that an inflaton mass-squared term is the only relevant 
SUGRA correction during inflation in a scenario with 
Planck-scale inflaton field values seems totally 
unjustified. In addition, this paper follows the usual 
practice of only taking into account the radiative 
corrections in the derivatives of the inflationary 
potential when calculating the slow-roll parameters and 
neglecting them when calculating the potential itself. In 
the case of extremely strong radiative corrections, 
however, this may lead to erroneous results. Also, 
Ref.~\cite{ck} attempts to accommodate the BICEP2 results 
in a double hybrid inflation model where the inflaton 
potential changes dynamically with the evolution of the 
inflaton fields. The particular implementation of this 
interesting idea, though, appears to have some problems 
of naturalness in the design of the superpotential. In 
addition, the treatment of SUGRA seems to be incomplete.   
Finally, Ref.~\cite{rsw} shows that, in 
SUSY hybrid inflation models, it is possible to obtain 
values of $r$ close to $0.03$ by employing an expansion 
of a non-minimal K\"{a}hler potential with appropriate 
coefficients. The validity of this approach may, however,
be questionable since the inflaton takes values close to
the Planck scale.   

In this paper, we will show that a reduced version of the 
extended SUSY PS model of Ref.~\cite{quasi} can yield a 
two stage inflationary scenario which can predict values of 
$r$ up to about $0.05$. Larger values of the tensor-to-scalar 
ratio would lead to unacceptably large running of the scalar 
spectral index. In the range of the model parameters 
considered here, the model in the global SUSY limit possesses 
practically two classically flat directions, namely the 
trivial and the semi-shifted \cite{semi} one. After including 
SUGRA corrections, the trivial path, on which the full GUT 
gauge group is unbroken, is stabilized and a first stage of 
inflation 
can occur as the system slowly rolls down this path. All the 
cosmological scales exit the horizon during this stage and 
our present horizon undergoes a limited number of e-foldings.
The obtained tensor-to-scalar ratio can be appreciable while 
the scalar spectral index assumes acceptable values thanks 
to the competing effect of the sufficiently mild SUGRA
corrections resulting from the construction of 
Ref.~\cite{pana} and the strong radiative corrections to the 
inflationary potential. 

Subsequently, a second inflationary stage occurs along 
the semi-shifted path, where $U(1)_{B-L}$ remains unbroken, 
and provides the additional number of e-foldings 
required for solving the standard problems of hot big bang 
cosmology. This is possible since, for our 
choice of parameters, the semi-shifted path is almost 
perpendicular to the trivial path and, thus, is not 
affected by the strong radiative corrections on the 
trivial path. It is also important that the SUGRA corrections 
on the semi-shifted path are kept sufficiently mild again by 
the mechanism of Ref.~\cite{pana}.

We first present, in Sec.~\ref{sec:susy}, the salient 
features of the model in global SUSY. In Sec.~\ref{sec:sugra}, 
we then calculate the SUGRA and one-loop radiative 
corrections to the potential and discuss our double 
inflationary scenario. Finally, in Sec.~\ref{sec:concl}, we 
summarize our conclusions. Throughout, we will use units where 
the reduced Planck mass $m_{\rm P}=2.4354\times 10^{18}~{\rm GeV}$ 
is equal to unity.                 

\section{The model in global SUSY} 
\label{sec:susy}

We consider a reduced version of the extended SUSY PS 
model of Ref.~\cite{quasi}. This version is based on the 
left-right symmetric gauge group $G_{\rm LR}=SU(3)_c\times 
SU(2)_{\rm L}\times SU(2)_{\rm R}\times U(1)_{B-L}$, which 
is a subgroup of the PS group. The superfields of the model 
which are relevant for inflation are the following. A gauge 
singlet $S$, a pair of superfields $\Phi$, 
$\bar{\Phi}$ belonging to the $(1,1,3)_{0}$ representation 
of $G_{\rm LR}$, and a conjugate pair of Higgs superfields 
$H$ and $\bar{H}$ belonging to the $(1,1,2)_{1}$ and 
$(1,1,2)_{-1}$ representations of $G_{\rm LR}$, respectively.  
The field $\Phi$ acquires a vacuum expectation value (VEV) 
which breaks $G_{\rm LR}$ to $G_{\rm SM}\times U(1)_{B-L}$ 
with $G_{\rm SM}$ being the standard model (SM) gauge group,
while the VEVs of $H$ and $\bar{H}$ cause the breaking of 
$G_{\rm LR}$ to $G_{\rm SM}$. The full superfield content and 
superpotential, 
the global symmetries, and the charge assignments can be easily 
derived from the extended SUSY PS model of Ref.~\cite{quasi} by 
simply reducing its GUT gauge group to $G_{\rm LR}$. The only 
global symmetry of the model which is relevant here is its 
$U(1)$ R symmetry under which $S$ and $\bar\Phi$ have charge 1 
with all the other superfields mentioned above being neutral.

The superpotential terms relevant for inflation are
\beq
\label{eq:superpotential}
W=\kappa S\left(M^2-\Phi^2\right)-\gamma SH\bar{H}+
m\Phi \bar{\Phi} -\lambda\bar{\Phi}H\bar{H},
\eeq
where $M$, $m$ are superheavy masses
and $\ka$, $\gamma$, $\la$ are dimensionless coupling constants. 
These parameters are normalized so that they correspond to the 
couplings between the SM singlet components of the superfields.
The mass parameters $M$, $m$ and any two of the three 
dimensionless parameters $\ka$, $\gamma$, $\la$ can always be 
made real and positive by appropriately redefining the phases 
of the superfields. The third dimensionless parameter, however, 
remains generally complex. For definiteness, we will choose 
this parameter to be real and positive too.

The F--term scalar potential obtained from the superpotential 
in Eq.~(\ref{eq:superpotential}) is given by
\bea
\label{eq:SUSYpotential}
V^0_F&=&|\kappa(M^2-\Phi^2)-\gamma H\Hb|^2
\nonumber\\
& &+|m\Pb-2\ka S\Phi|^2+|m\Phi-\la H\Hb|^2
\nonumber\\
& &+|\gamma S+\la\Pb\,|^2\left(|H|^2+|\Hb|^2
\right),
\eea
where the complex scalar fields which belong to
the SM singlet components of the superfields are
denoted by the same symbol. From this potential 
and the vanishing of the D--terms (which implies 
that $\bar{H}^{*}=e^{i\theta}H$), one finds
\cite{semi} two distinct continua of SUSY vacua:
\bea
\Phi=\Phi_{+},\,\,\bar{H}^{*}=H,\,\, |H|=
\sqrt{\frac{m\Phi_{+}}{\la}}, \,\, S=\bar\Phi=0,
\label{eq:vacua+}\\
\Phi=\Phi_{-},\,\,\bar{H}^{*}=-H,\,\, |H|=
\sqrt{\frac{-m\Phi_{-}}{\la}}, \,\, S=\bar\Phi=0,
\label{eq:vacua-}
\eea
where
\begin{equation}
\Phi_{\pm}\equiv\pm M\left(\sqrt{1+\left(\frac{\gamma m}{2\ka\la M}
\right)^2}\mp\frac{\gamma m}{2\ka\la M}\right).
\label{eq:SUSYvacua1}
\end{equation} 

The potential in Eq.~\eqref{eq:SUSYpotential}, generally, 
possesses \cite{semi} three flat directions. The first one 
is the usual trivial flat direction at
\begin{equation}
\Phi=\Pb=H=\Hb=0
\end{equation}
with
\begin{equation}
V^0_F=V_\text{tr}\equiv\ka^2M^4.
\end{equation}
On this direction, $G_{\rm LR}$ is unbroken. The second 
one, which appears at
\begin{gather}
\Phi=-\frac{\gamma m}{2\ka\la},\quad \Pb=-\frac{\gamma}{\la}\,
S,
\nonumber \\
H\Hb=\frac{\ka\gamma(M^2-\Phi^2)+\la m\Phi}
{\gamma^2+\la^2},
\nonumber \\
V^0_F=V_\text{nsh}\equiv\ka^2M^4\left(\frac{\la^2}
{\gamma^2+\la^2}\right)\left(1+\frac{\gamma^2m^2}
{4\ka^2\la^2M^2}\right)^2,
\end{gather}
is the trajectory for the new shifted hybrid inflation 
\cite{nshift}. On this direction, $G_{\rm LR}$ is broken to 
$G_{\rm SM}$. The third flat direction, which exists only 
if $M^2>m^2/2\ka^2$, lies at
\beq
\Phi=\pm\,M\sqrt{1-\frac{m^2}{2\kappa^2M^2}},\quad
\Pb=\frac{2\ka\Phi}{m}\,S,\quad H=\Hb=0.
\label{eq:semishift}
\eeq
It is the path along which semi-shifted hybrid inflation 
\cite{semi} takes place with
\beq
V^0_F=V_\text{ssh}\equiv
m^2M^2\left(1-\frac{m^2}{4\ka^2M^2}\right).
\eeq
Along this direction $G_{\rm LR}$ is broken to
$G_{\rm SM}\times U(1)_{B-L}$.

We choose to consider the case where $M^2>m^2/2\ka^2$
and, thus, the semi-shifted flat direction exists. One can 
show -- see Ref.~\cite{semi} -- that, in this case, we always 
have $V_\text{ssh}<V_\text{nsh}$ and $V_\text{ssh}<V_\text{tr}$.
Therefore, the semi-shifted flat direction, if it exists, 
always lies 
lower than both the trivial and the new shifted one. On the 
other hand, the new shifted flat direction may either lie lower 
or higher than the trivial one depending on the values of the 
parameters. Here we will take $\ka\sim 1$, $\gamma\ll \la\ll
\kappa$, $m\ll M$, and $|S|<1$. In this case, the new shifted flat 
direction practically coincides with the trivial one and, thus, 
plays no independent role in our scheme. 

\section{The double inflationary scenario}
\label{sec:sugra}

In this section, we will show that, after including SUGRA 
corrections, the trivial path becomes stable for large 
absolute values of the real canonically normalized inflaton. 
Thus, it can support a first stage of inflation during which 
the universe undergoes a number of e-foldings which, although 
limited, is adequately large for all the cosmological scales 
to exit the horizon. Strong radiative corrections to the 
inflationary potential, which are controlled by the parameter 
$\ka$, in conjunction with mild SURGA corrections then 
guarantee that an appreciable value of the tensor-to-scalar 
ratio can be achieved together with an acceptable value of the 
scalar spectral index. Our scenario can predict values of 
the tensor-to-scalar ratio only up to about $0.05$ because 
larger values require unacceptably large running of the 
scalar spectral index. 

A subsequent second stage of inflation along the semi-shifted 
path can provide the additional number of e-foldings 
required for solving the horizon and flatness problems of the
standard hot big bang cosmology. This is possible since, for 
the parameters chosen, this direction is almost orthogonal to 
the trivial path and, thus, it is not affected by the 
strong radiative corrections present during the first 
stage of inflation. In this connection, it is also important 
that the SUGRA corrections on the semi-shifted path remain 
mild. 

After the termination of the first stage of inflation, the 
system moves towards the semi-shifted path and the group 
$SU(2)_{\rm R}$ breaks spontaneously to a $U(1)$ subgroup,
leading to the formation of magnetic monopoles. On the other 
hand, the spontaneous breaking of a linear combination of 
this $U(1)$ and $U(1)_{B-L}$, which takes 
place at the end of the second inflationary stage, leads to 
the production of open cosmic strings which connect these 
monopoles to antimonopoles. Subsequently, the monopoles come 
into the post-inflationary horizon and the whole system of 
strings and monopoles decays well before recombination without 
leaving any trace in the CMBR. The gravitational waves which 
are generated by the decaying strings may, though, be 
measurable in the future.   
              
\subsection{The first inflationary stage}

We adopt here the following K\"{a}hler potential
\begin{eqnarray}
\label{kaehler}
K&=&-\ln\left(1-|S|^2\right)-\ln\left(1-|\bar{\Phi}|^2\right)+
|\Phi|^2+|H|^2 \nonumber \\
& &+|\bar H|^2-2\ln\left(-\ln|Z_1|^2\right)+|Z_2|^2
\end{eqnarray}
($|S|,\,|\bar{\Phi}|<1,\,0<|Z_1|<1$), where we included two 
extra $G_{\rm LR}$ singlet superfields 
$Z_1$ and $Z_2$, which do not enter the superpotential at all 
because
they transform non-trivially under additional anomalous $U(1)$ 
gauge symmetries. The resulting F--term potential in SUGRA is 
given by
\begin{equation}
V_F=\left[\sum_i |W_{X_i}+K_{X_i}W|^2K_{X_i{X_i}^*}^{-1}-3|W|^2 
\right]e^K,
\label{sugrapot}
\end{equation}
where a subscript $X_i$ denotes derivation with respect to the 
field $X_i$ 
and the sum extends over all the seven fields $S,\,\bar{\Phi},\,
\Phi,\,H,\,\bar{H},\,Z_1,\,Z_2$. The values of $Z_1$ and $Z_2$ 
are assumed to be fixed \cite{pana} by anomalous D--terms. Note 
that the superfields $S,\,\bar{\Phi},\,Z_1$ possess K\"{a}hler 
potentials of the no-scale type which for $Z_2=0$, in view of 
the relation
\begin{equation}
|K_{Z_1}|^2K_{Z_1{Z_1}^*}^{-1}=2,
\end{equation}
guarantee the exact flatness of the potential along 
the trivial path \cite{pana} and its approximate flatness on 
the semi-shifted one -- see below. These paths are, 
respectively, parametrized by the complex inflatons $S$ and 
$\bar\Phi$ 
(approximately). However, as we shall see -- cf. Ref.~\cite{pana} 
-- , the relation  
\begin{equation}
|K_{Z_2}|^2K_{Z_2{Z_2}^*}^{-1}=|Z_2|^2\equiv\beta
\end{equation}
implies that these inflatons acquire squared masses proportional 
to $\beta$ as soon as the value of $Z_2$ becomes non-zero.

Using the R and $U(1)_{B-L}$ symmetries of the model, we can rotate
$S$ and $H$ on the real axis -- cf. e.g. Ref.~\cite{semi}. The
fields $\Phi,\,\bar{\Phi},\,\bar{H}$ remain in general complex. 
However, for simplicity, we will also restrict them on the real 
axis. This is not expected to influence our results in any 
essential way since these fields are anyway real in the vacuum
and on all the flat directions of the model given that the 
parameters of the model are chosen to be real -- see 
Sec.~\ref{sec:susy}. Also, we can show that, everywhere on the 
trivial and the semi-shifted inflationary paths, the mass 
squared matrices of the imaginary parts of the fields do not mix 
with the mass squared matrices of their real parts and, during 
both inflations, have positive eigenvalues in the directions 
perpendicular to these paths. So there is no instability in the 
direction of the imaginary parts of the fields which are 
orthogonal to these inflationary trajectories. Moreover, as we 
can demonstrate, both the trivial and the semi-shifted 
inflationary paths are destabilized with the fields developing 
real values. 

We define the canonically normalized real scalar 
fields $\sigma,\,\bar\phi,\,\phi,\,h,\,\bar{h}$ corresponding 
to the K\"{a}hler potential in Eq.~(\ref{kaehler}) as follows 
-- cf. Ref.~\cite{pana} -- : 
\begin{equation}
S=\tanh\frac{\sigma}{\sqrt{2}}, \quad \bar{\Phi}=
\tanh\frac{\bar{\phi}}{\sqrt{2}}, 
\end{equation}
\begin{equation}
\Phi=\frac{\phi}{\sqrt{2}}, \quad H=\frac{h}{\sqrt{2}}, 
\quad\bar{H}=\frac{\bar{h}}{\sqrt{2}}.
\end{equation}
 
We can now evaluate the potential $V_F$ in Eq.~(\ref{sugrapot}) 
with the overall factor $\exp{\left[-2\ln\left(-\ln|Z_1|^2\right)+
|Z_2|^2\right]}$ 
absorbed into redefined parameters $\kappa$, $\gamma$, $m$, and 
$\lambda$ and find
\begin{eqnarray}
 V_F  & = & \left[A_1^2 \cosh^2 \frac{\bar{\phi}}{\sqrt{2}} 
-A_2^2 \sinh^2 \frac{\bar{\phi}}{\sqrt{2}} +\beta A_3^2 +A_4^2+A_5^2 
\right. \nonumber \\
& &+  \left. \frac{1}{2} \left(h^2+\bar{h}^2 \right)A_6^2+ 
\frac{1}{2} \left(\phi^2+ h^2+\bar{h}^2 \right)A_3^2 \right. 
\nonumber \\ 
& & + \left. \left(\sqrt{2} \phi A_5-2h\bar{h}A_6\right)A_3 
\right]e^{\frac{1}{2}\left(\phi^2+ h^2+\bar{h}^2 \right)} .
\label{pot}
\end{eqnarray}
Here   
\begin{equation}
\label{a1}
A_1=\kappa\left(M^2-\frac{\phi^2}{2}\right)-\frac{\gamma}{2} 
h\bar{h},
\end{equation}
\begin{equation}
A_2=m\frac{\phi}{\sqrt{2}}- \frac{\lambda}{2} h\bar{h},
\end{equation}
\begin{equation}
A_3=A_1 \sinh \frac{\sigma}{\sqrt{2}} \cosh \frac{\bar{\phi}}
{\sqrt{2}} + A_2 \cosh \frac{\sigma}{\sqrt{2}} 
\sinh \frac{\bar{\phi}}{\sqrt{2}}, 
\end{equation}
\begin{equation}
A_4=A_1 \sinh \frac{\sigma}{\sqrt{2}} \sinh \frac{\bar{\phi}}
{\sqrt{2}} + A_2 \cosh \frac{\sigma}{\sqrt{2}} \cosh 
\frac{\bar{\phi}}{\sqrt{2}}, 
\end{equation}
\begin{equation}
A_5=m \cosh \frac{\sigma}{\sqrt{2}} \sinh \frac{\bar{\phi}}
{\sqrt{2}} - {\sqrt{2}} \kappa \phi \sinh \frac{\sigma}
{\sqrt{2}} \cosh \frac{\bar{\phi}}{\sqrt{2}}, 
\end{equation}
and
\begin{equation}
A_6=\gamma \sinh \frac{\sigma}{\sqrt{2}} \cosh \frac{\bar{\phi}}
{\sqrt{2}} + \lambda  \cosh \frac{\sigma}{\sqrt{2}} \sinh 
\frac{\bar{\phi}}{\sqrt{2}}. 
\end{equation}

On the trivial trajectory where $\bar{\phi},\,\phi,\,h,\,
\bar{h} =0$, the F--term potential takes the form
\begin{equation}
\label{eq:VF1}
V_F=\kappa^2 M^4\left[1+\beta \sinh^2 \frac{\sigma}
{\sqrt{2}}\right].
\end{equation}
The mass-squared eigenvalues in the directions perpendicular to 
this trajectory for $\sinh^2\left(\sigma/\sqrt{2}\right)\gg M^2/2$
can also be found from Eq.~(\ref{pot})
to be 
\begin{equation} 
m_{\phi}^2\simeq 4\kappa^2 \sinh^2\frac{\sigma}{\sqrt{2}},
\end{equation} 
\begin{equation} 
m_{\bar{\phi}}^2\simeq \kappa^2M^4\left(1+(1+\beta)\sinh^2
\frac{\sigma}{\sqrt{2}}\right),
\end{equation} 
\begin{equation}
\label{chi1} 
m^2_{{\chi}_1}\simeq(\kappa M^2-\gamma)\left[\kappa M^2+
\left((1+\beta)\kappa M^2-\gamma\right)\sinh^2\frac{\sigma}
{\sqrt{2}}\right],
\end{equation} 
and
\begin{equation} 
\label{chi2}
m^2_{{\chi}_2}\simeq(\kappa M^2+\gamma)\left[\kappa M^2+
\left((1+\beta)\kappa M^2+\gamma\right)\sinh^2\frac{\sigma}
{\sqrt{2}}\right],
\end{equation} 
where $\chi_1=(h+\bar{h})/\sqrt{2}$ and $\chi_2=(h-\bar{h})/
\sqrt{2}$. Thus, assuming that $\gamma < \kappa M^2$, we see 
that the trivial path, which is flat in the limit $\beta \to 
0$, is stable for large absolute values of the inflaton 
$\sigma$. 

Note that Eqs.~(\ref{chi1}) and (\ref{chi2}) hold for any 
value of $\sinh^2\left(\sigma/\sqrt{2}\right)$. On the 
contrary, one can show that as $\sinh^2\left(\sigma/\sqrt{2}
\right)$ decreases, the eigenvalues and eigenstates of the 
mass-squared matrix of the $\phi-\bar{\phi}$ system change. 
In particular, when $\sinh^2\left(\sigma/\sqrt{2}\right)
\simeq M^2/2+m^2/(2\kappa^2 M^2)$, the mass-squared matrix of 
the $\phi-\bar{\phi}$ system acquires a zero eigenvalue with 
$\bar{\phi}$ dominating the corresponding eigenstate. 
Subsequently, as $\sinh^2\left(
\sigma/\sqrt{2}\right)$ approaches the value $M^2/2$, the 
eigenvalues of the $\phi-\bar{\phi}$ mass-squared matrix 
become almost opposite to each other with $\phi$ and 
$\bar{\phi}$ contributing almost equally to both the 
eigenstates. A further decrease of $\sinh^2\left(\sigma/
\sqrt{2}\right)$ leads to the domination of the unstable 
eigenstate by $\phi$. Since the field $\phi$ is required 
to develop a nonzero VEV in order to cancel the false 
vacuum energy density $\kappa^2 M^4$ on the trivial 
trajectory -- see Eq.~(\ref{eq:SUSYpotential}) or 
Eqs.~(\ref{pot}) and (\ref{a1}) -- , we will take as  
critical value $\sigma_{\rm c}$ of $\sigma$ at which the 
trivial path is destabilized the one determined by the 
relation
\begin{equation}
\label{critical1}
\sinh^2\frac{\sigma_{\rm c}}{\sqrt{2}}=\frac{M^2}{2}.
\end{equation}

To the F--term scalar potential $V_F$ in Eq.~(\ref{pot}) 
has to be added during the first stage of inflation (i.e. 
for $|\sigma|\ge |\sigma_{\rm c}|$) the term
\begin{equation}
V_r^{\phi}=\kappa^2 M^4 \left(\frac{N_{\phi}\kappa^2}
{8\pi^2}\right) \ln \frac{2 \tanh^2\frac{\sigma}
{\sqrt{2}}}{M^2}
\end{equation}
corresponding to the dominant one-loop radiative 
corrections to the inflationary potential due to the 
$N_{\phi}$-dimensional supermultiplet $\Phi$ ($N_{\phi}=3$). 
Notice that the renormalization scale in these radiative 
corrections is chosen such that  $V_r^{\phi}$ vanishes at 
$|\sigma|=|\sigma_{\rm c}|$ ($\tanh^2\left(\sigma_{\rm c}/
\sqrt{2}\right)\simeq\sinh^2\left(\sigma_{\rm c}/\sqrt{2}
\right)=M^2/2$).

Setting
\begin{equation}
 \delta_{\phi}=\frac{N_{\phi}\kappa^2}{2\pi^2},
\end{equation}
we can rewrite the full inflationary potential and its 
derivatives (denoted by primes) with respect to the 
canonically normalized real inflaton field $\sigma$ as 
follows:
\begin{equation}
\frac{V}{\kappa^2 M^4}=1+\beta \sinh^2 \frac{\sigma}{\sqrt{2}}
+\frac{\delta_{\phi}}{4} \ln \frac{2 \tanh^2\frac{\sigma}
{\sqrt{2}}}{M^2} \equiv C(\sigma),
\end{equation}
\begin{equation}
\label{Vprime}
\frac{V^{\prime}}{\kappa^2 M^4}=\frac{1}{\sqrt{2}}
\sinh(\sqrt{2}\sigma)\left(\beta+\frac{\delta_{\phi}}
{\sinh^2(\sqrt{2}\sigma)}\right),
\end{equation}
\begin{equation}
\label{Vprimeprime}
\frac{V^{\prime \prime}}{\kappa^2 M^4}= \cosh(\sqrt{2}\sigma)
\left(\beta-\frac{\delta_{\phi}}{\sinh^2(\sqrt{2}\sigma)}\right),
\end{equation}
and
\begin{eqnarray}
\frac{V^{\prime \prime \prime}}{\kappa^2 M^4}&=& \sqrt{2}
\sinh(\sqrt{2}\sigma)\left(\beta-\frac{\delta_{\phi}}
{\sinh^2(\sqrt{2}\sigma)}\right) \nonumber \\ 
& & +\frac{2\sqrt{2}\delta_{\phi}}{\tanh^2(\sqrt{2}\sigma)
\sinh(\sqrt{2}\sigma)}.
\end{eqnarray}
The usual slow-roll parameters for inflation are then written as
\begin{equation}
\label{epsilon}
\epsilon=\frac{1}{2}\left(\frac{V^{\prime}}{\kappa^2 M^4}
\right)^2\frac{1}{C^2(\sigma)},
\end{equation}
\begin{equation}
\eta=\left(\frac{V^{\prime \prime}}{\kappa^2 M^4}\right)\frac{1}
{C(\sigma)},
\end{equation}
and
\begin{eqnarray}
\xi&=&\left(\frac{V^{\prime}}{\kappa^2 M^4}\right)\left(
\frac {V^{\prime \prime \prime}}{\kappa^2 M^4}\right)
\frac{1}{C^2(\sigma)}=2\left|\tanh(\sqrt{2}\sigma)\right|
\eta\sqrt{\epsilon}
\nonumber \\  
& &+\frac{4\delta_{\phi}\sqrt{\epsilon}}{C(\sigma)
\tanh^2(\sqrt{2}\sigma)\left|\sinh(\sqrt{2}\sigma)\right|}. 
\end{eqnarray}
Using these expressions, we can evaluate the scalar 
spectral index $n_{\rm s}$, its running $\alpha_{\rm s}$, 
and the tensor-to-scalar ratio $r$ from the formulas
\begin{equation}
\label{ns}
n_{\rm s}=1+2\eta-6\epsilon,
\end{equation}
\begin{equation}
\alpha_{\rm s}=16\eta\epsilon-24\epsilon^2-2\xi,
\end{equation}
\begin{equation}
\label{ratio}
r=16\epsilon.
\end{equation}
Finally, the scalar potential on the trivial inflationary 
path can be written in terms of the scalar power spectrum 
amplitude $A_{\rm s}$ and $r$ as follows
\begin{equation}
V=\frac{3\pi^2}{2}A_{\rm s} r.
\end{equation}

As a numerical example, we take the value of the real inflaton 
field $\sigma$ at horizon exit of the pivot scale 
$k_*=0.05 \ \rm {Mpc}^{-1}$ to be $\sigma_*=1.45$. Also, we 
take $\kappa=1.7$, $\beta=0.022$, and the scalar power spectrum 
amplitude $A_{\rm s}=2.215 \times 10^{-9}$ at the 
same pivot scale \cite{planck15}. With these input numbers, we 
then find $M=3.493 \times 10^{-3}$, $C(\sigma_*)=2.2941$, 
$\epsilon=0.00188$, $\eta=-0.01389$, $n_{\rm s}=0.9609$, 
$\alpha_{\rm s}=-0.01674$, and $r=0.0301$. 

As one can see, our predictions can not only be perfectly 
consistent with the latest data released by the Planck 
satellite experiment \cite{planck15}, but also accommodate large 
values of the tensor-to-scalar ratio $r$ of order ${\rm few} 
\times 10^{-2}$. As it is obvious from Eq.~(\ref{ratio}), such 
values of $r$ require relatively large values of $\epsilon$, 
which in turn reduce the scalar spectral index $n_{\rm s}$ in 
Eq.~(\ref{ns}) below unity, but not quite adequately to make it 
compatible with the data. So we need an appreciable negative 
value of $\eta$, which requires that the parenthesis in the 
right hand side of Eq.~(\ref{Vprimeprime}) be dominated by the 
second term. A similar parenthesis appears
in the right hand side of Eq.~(\ref{Vprime}) too, but with the 
two terms added rather than subtracted. As it turns out, both 
these terms have to be appreciable with the second one being 
larger in order to be able to bring $n_{\rm s}$ near its 
best-fit value from the Planck data. This is certainly possible 
only for large values of the parameter $\kappa$ controlling the 
radiative corrections on the trivial path. Note that the first 
stage of inflation ends before the system reaches the critical 
point in Eq.~(\ref{critical1}) by violating the slow-roll 
conditions and the obtained number of e-foldings is limited due 
to the large values of $\epsilon$ involved and the fact that 
$\sigma_*\sim 1$. 

\subsection{The second inflationary stage}

For the rest of the parameters of the model, we chose 
the values $m=1.827 \times 10^{-5}$, $\lambda=0.1$, 
and $\gamma=10^{-6}$. We solved numerically the 
differential equations of the system with potential 
energy density given by the exact $V_F$ in Eq.~(\ref{pot})
supplemented with the relevant radiative corrections and 
the D--terms involving the fields $H$,
$\bar{H}$. The numerical investigation then revealed 
that there exist appropriate small initial absolute values 
of the scalar fields $\bar{\phi},\, \phi,\, h,\, \bar{h}$ 
for which, after the first stage of inflation and the 
elapse of a sufficient amount of cosmic time $t$ for the 
energy density to approach $m^2M^2$, we have 
${\phi^2}\simeq 2M^2$, $h, \,\bar{h} \simeq 0$, and the scalar 
fields $\sigma$ and $\bar{\phi}$ take values such that $A_5 
\simeq 0$ with $|\sigma| \ll 1$. So it is obvious that the 
system reaches the semi-shifted inflationary path in 
Eq.~(\ref{eq:semishift}) -- 
note that the second relation in this equation is equivalent 
to $A_5=0$. It is remarkable that $|\bar\phi|$, which at the 
end of the first inflationary stage is extremely small, 
manages to attain values of the order of ${\rm few}\times 
10^{-1}$ at the onset of the second stage. 

It is worth noticing, however, that the initial values of the 
fields which lead to a double inflation scenario, although 
quite frequent, do not seem to form well-defined connected 
regions. In other words, the solutions of the coupled system 
of differential equations exhibit a rather `chaotic' behavior 
in the sense that a slight change of the initial conditions 
may lead from a double to a single inflation scenario. Note 
that a similar situation is encountered \cite{initial} even 
in the simplest SUSY hybrid inflation scenario, where a 
slight change of initial conditions may ruin inflation 
leading the system directly to the vacuum. Such a behavior, 
given the multidimensionality of the field space in our 
case, makes it very difficult to provide further details 
concerning the structure of the space of initial values 
leading to the desirable scenario.  
 
For a negligible value of $\gamma$, ${\phi^2}\simeq 2M^2 $, 
$A_5\simeq 0$,
and $|\sigma|\ll 1$, we find that $A_1\simeq 0$, $A_3\simeq A_2 
\sinh\left(\bar{\phi}/\sqrt{2}\right)$, $A_4 \simeq A_2\cosh 
\left(\bar{\phi}/\sqrt{2}\right)$, $A_6\simeq\lambda\sinh\left(
\bar{\phi}/\sqrt{2}\right)$, and the F--term scalar potential 
becomes
\begin{eqnarray}
\label{secondpot}
V_F  & \simeq &\left[A_2^2  +(\beta +M^2) A_2^2 \sinh^2 
\frac{\bar{\phi}}{\sqrt{2}} \right. \nonumber \\ & & 
+\left. \frac{1}{2}\left(h^2+\bar{h}^2 \right)\left(
\lambda^2+A_2^2 \right)\sinh^2 \frac{\bar{\phi}}{\sqrt{2}} 
\right. \nonumber \\
& & \left.-2h\bar{h}\lambda A_2\sinh^2 \frac{\bar{\phi}}
{\sqrt{2}}\right]e^{M^2+\frac{1}{2}\left( h^2+\bar{h}^2 
\right)} \\
& \overset{h,\bar{h} \simeq 0}{\underset{M^2\ll \beta}{\simeq}} 
& m^2M^2\left[1+\beta \sinh^2 \frac{\bar{\phi}}{\sqrt{2}}\right]. 
\label{VF2}
\end{eqnarray}
The expression in Eq.~(\ref{VF2}) gives approximately 
the F--term potential on the semi-shifted path. Notice the 
striking similarity of this expression with the one in 
Eq.~(\ref{eq:VF1}) involving the same parameter $\beta$. 

From $A_5 \simeq 0$ and the fact that $A_5 \propto m\tanh\left(
\bar{\phi}/\sqrt{2}\right)-\sqrt{2}\kappa \phi \tanh \left(
\sigma/\sqrt{2}\right)$, it follows that the combination of $S$ 
and $\bar{\Phi}$ which could remain large when the energy 
density approaches $ m^2M^2$ and plays the role of the complex 
inflaton in the second stage of inflation is 
\beq
\frac{mS+2\kappa <\Phi>\bar{\Phi}}
{\sqrt{m^2+4 \kappa^2 M^2}}\simeq \bar{\Phi},
\eeq
since the contribution of $\bar{\Phi}$ in this combination
is about $2 \kappa M /m \simeq 650$ times bigger than the one of 
$S$.

From Eq.~(\ref{secondpot}), we can construct the mass-squared 
matrix for the $h-\bar{h}$ system during the second stage of 
inflation. We find that the mass eigenstates are given by the 
combinations $\chi_1=(h+\bar{h})/\sqrt{2}$ and $\chi_2=
(h-\bar{h})/\sqrt{2}$ with masses-squared 
\beq
m^2_{{\chi}_1}\simeq\left(\lambda-mM\right) \left[\left( 
\lambda -(1+\beta)mM\right)  \sinh^2 \frac{\bar{\phi}}
{\sqrt{2}}-mM\right],
\eeq
\beq
m^2_{{\chi}_2}\simeq\left(\lambda+mM\right) \left[\left( 
\lambda +(1+\beta)mM\right)  \sinh^2 \frac{\bar{\phi}}
{\sqrt{2}}+mM\right].
\eeq
We see that ${\chi}_1$ develops an instability which terminates 
the valley along which the second stage of inflation takes 
place with the critical value $\bar{\phi}_{\rm c}$ of the real 
canonically normalized inflaton $\bar{\phi}$ being approximately 
determined from the relation
\beq
\sinh^2 \frac{\bar{\phi}_{\rm c}}{\sqrt{2}} = \frac{mM}{\lambda}.
\eeq

To the F--term scalar potential $V_F$ during the second stage of 
inflation (i.e. for $|\bar{\phi}| \ge |\bar{\phi}_{\rm c}|$ 
and $|\sigma|<|\sigma_{\rm c}|$ ) 
has to be added the following term 
\begin{eqnarray}
V_r^{h}&=& m^2 M^2 \left(\frac{N_h \lambda^2}{16\pi^2}\right) 
\nonumber \\
& &\times\ln \frac{\left(\tanh\frac{\sigma}{\sqrt{2}}+\sqrt{2}
\kappa 
\frac{<\phi>}{m}\tanh \frac{\bar{\phi}}{\sqrt{2}}\right)^2}
{\left(1+4 \kappa^2 \frac{M^2}{m^2} \right)\left(\frac{mM}
{\lambda}\right)},
\end{eqnarray}
which may be approximated as
\beq
\label{eq:Vrh}
V_r^{h} \simeq m^2 M^2 \left(\frac{N_{h}\lambda^2}{16\pi^2}
\right) \ln \frac{\lambda \tanh^2 \frac{\bar{\phi}}
{\sqrt{2}}}{mM}
\eeq
and corresponds to the dominant one-loop radiative corrections due 
to the $N_{h}$-dimensional supermultiplets $H$, $\bar{H}$ 
($N_{h}=2$). Notice that the renormalization scale is chosen such 
that  $V_r^{h}$ vanishes at $|\bar{\phi}|=|\bar{\phi}_{\rm c}|$ 
($\tanh^2\left(\bar{\phi}_{\rm c}/\sqrt{2}\right)\simeq 
\sinh^2\left(\bar{\phi}_{\rm c}/\sqrt{2}\right)=mM/\lambda$). 

The one-loop radiative corrections involving the $\Phi$ 
supermultiplet are neglected since they are relatively very small. 
This is because $\Phi$ couples to the combination which plays the 
role of the complex inflaton during the second stage of inflation 
only through $S$ and the contribution of $S$ to this combination is 
severely suppressed. Indeed, the slope of the potential along the 
semi-shifted path generated by the radiative corrections involving 
the $\Phi$ supermultiplet is suppressed relative to the one 
involving the $H$, $\bar{H}$ supermultiplets by, approximately, a 
factor $\left(N_{\phi}/8N_{h}\right)\left(m/\lambda M\right)^2\sim 
5\times 10^{-4}$. This is a very important property of our model
resulting from the fact that, for the parameters chosen, the 
semi-shifted path is almost perpendicular to the trivial one. So 
the very strong radiative corrections on the trivial trajectory, 
which are controlled by the strong coupling constant $\kappa$ and 
are needed, as we have seen, for accommodating appreciable values 
of $r$, do not affect the second stage of inflation. This is very 
crucial since otherwise the semi-shifted path would become too 
steep and there would be no way of generating the extra e-foldings 
required for solving the puzzles of hot big bang cosmology.      

The number of e-foldings during the second stage of inflation 
between an initial value $\bar{\phi}_{\rm {in}}$ and a final 
value $\bar{\phi}_{\rm {f}}$ of the inflaton $\bar{\phi}$ is 
given, in the slow-roll approximation, by $N(\bar{\phi}_{\rm {f}})
-N(\bar{\phi}_{\rm {in}})$, where
\beq
N(\bar{\phi}) \simeq \frac{1}{2\beta \sqrt{1-(\delta_{h}/\beta)}}
\ln \frac{\cosh(\sqrt{2} \bar{\phi})+\sqrt{1-(\delta_{h}/\beta)}}
{\cosh(\sqrt{2} \bar{\phi})-\sqrt{1-(\delta_{h}/\beta)}}
\label{ef}
\eeq
with 
\beq
\delta_{h}=\frac{N_{h}{\lambda^2}}{{4\pi^2}}.
\eeq 
The termination of
slow-roll inflation is due to the radiative corrections in 
Eq.~(\ref{eq:Vrh}) and takes place at a value $\bar{\phi}_{\rm f}$
($|\bar{\phi}_{\rm f}|\gg|\bar{\phi}_{\rm c}|$) of $\bar{\phi}$ 
given by
\beq
\cosh(\sqrt{2} \bar{\phi}_{\rm f}) \simeq 
\frac{\delta_{h}}{2}+\sqrt{1+\frac{\delta_{h}^2}{4}}.
\eeq

It turns out numerically that, with the chosen values of the 
parameters, the pivot scale $k_*=0.05~{\rm Mpc}^{-1}$ suffers 
about 13 e-foldings during the first stage of inflation. As a 
consequence, approximately another 38-39 e-foldings (for 
reheat temperature $T_{\rm r}=10^9~{\rm GeV}$) must be 
provided by the second inflationary stage, which requires a 
value of $|\bar{\phi}_{\rm in}| \simeq 0.23$ at the onset 
of this stage. This requirement can indeed be fulfilled in our 
numerical example as we have shown by extensive numerical 
studies. It is worth noticing that, due to the presence of mild
but appreciable SUGRA corrections and not too weak radiative 
corrections, the second stage of inflation is able to generate
a relatively limited number of e-foldings. Consequently, this 
number is not too sensitive to the value of 
$\bar{\phi}_{\rm in}$.

\begin{figure}[t]
\centerline{\epsfig{file=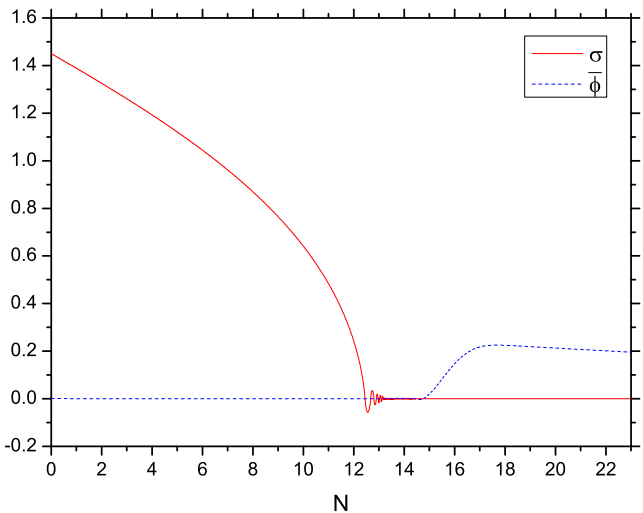,width=8.7cm}}
\caption{The evolution of the fields $\sigma$ and $\bar\phi$ 
for the case with $r=0.0301$ versus the number $N$ of 
e-foldings after the horizon exit of the pivot scale $k_*$, 
where we take $\sigma=1.45$, $\bar\phi=10^{-3}$, $\phi=10^{-8}$, 
$h=10^{-4}$, $\bar{h}=1.01\times 10^{-4}$, and $d\sigma/dt=
-1.1074\times 10^{-6}$.}
\label{fig1}
\end{figure}

In order to support the statements made above concerning the 
numerical part of our work, we depict, in Fig.~\ref{fig1}, 
the evolution of the fields $\sigma$ and $\bar\phi$
as functions of the number of e-foldings $N$ starting from the 
point where the pivot scale $k_*$ exits the horizon for a
particular choice of initial conditions at this point. Namely, 
we start with $\sigma=1.45$, $\bar\phi=10^{-3}$, $\phi=10^{-8}$, 
$h=10^{-4}$, 
and $\bar{h}=1.01\times 10^{-4}$ corresponding to an almost 
D-flat direction. All the fields are given zero initial velocity 
except for $\sigma$ the velocity of which is taken to be 
$-1.1074\times 10^{-6}$. This is, indeed, the actual value of 
the velocity of $\sigma$ on the trivial path, which was determined 
numerically.

We observe that $\sigma$ assumes values above its 
critical value for about $13$ e-foldings. Towards the end of the 
first inflationary stage, $\sigma$ oscillates with 
appreciable amplitude around zero four times. We may consider 
that, during these oscillations, the first stage of inflation has 
not come to an end. When the amplitude of the oscillations 
falls below the critical value of $\sigma$,  
$\phi$ moves to its value on the semi-shifted path and $\bar\phi$ 
starts performing slow oscillations with variable amplitudes 
typically of order $M$. The size of $\bar\phi$ remains 
small for about $1.7$ e-foldings before starting its growth and 
acquires its largest value $\simeq 0.225$ at $N\simeq 17.7$, when 
the second inflationary stage may be considered as having already 
started. For $N\gtrsim 20$, the evolution of $\bar\phi$ follows 
Eq.~(\ref{ef}) closely. 

\begin{figure}[t]
\centerline{\epsfig{file=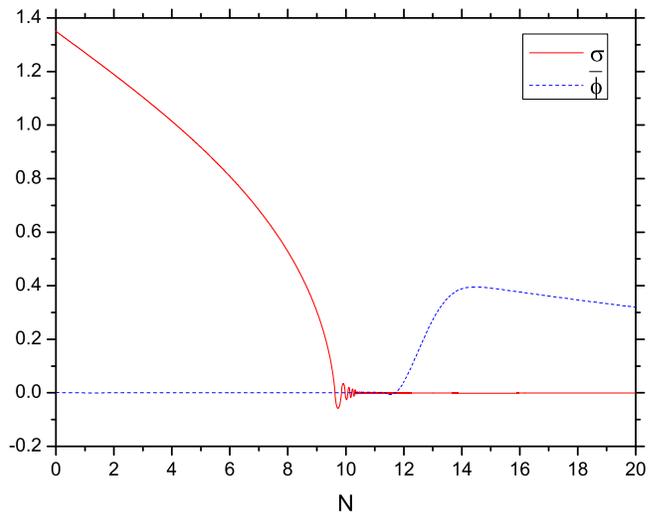,width=8.7cm}}
\caption{The evolution of the fields $\sigma$ and $\bar\phi$ for 
the case with $r=0.0502$ versus the number $N$ of e-foldings 
after the horizon exit of the pivot scale $k_*$, where we take 
$\sigma=1.35$, $\bar\phi=10^{-3}$, $\phi=10^{-8}$, $h=9\times 
10^{-4}$, $\bar{h}=9.01\times 10^{-4}$, and 
$d\sigma/dt=-1.8523\times 10^{-6}$.}
\label{fig2}
\end{figure}

If we allow for a stronger running of the scalar spectral index, 
we may obtain larger values of $r$. For example, taking 
$\sigma_*=1.35$, $\kappa=1.75$, and $\beta=0.037$, we find 
$M=3.891\times 10^{-3}$, $C(\sigma_*)=2.3479$, 
$\epsilon=0.00314$, $\eta=-0.00844$, $n_{\rm s}=0.9643$, 
$\alpha_{\rm s}=-0.03007$, and $r=0.0502$. In addition, we 
choose $m=3.891\times 10^{-5}$, $\lambda=0.1$, and 
$\gamma=10^{-6}$. With these choices, the pivot scale suffers 
about $10$ e-foldings during the first stage of inflation and, 
consequently, approximately another $41-42$ e-foldings must be 
provided by the second stage. This implies that 
$|\bar{\phi}_{\rm in}|$ lies in the range $0.38-0.40$. We 
verified numerically that the fulfillment of this requirement 
is indeed feasible. In Fig.~\ref{fig2}, we depict the evolution 
of the fields $\sigma$ and $\bar\phi$ as functions of the number 
of e-foldings $N$ again starting from the point where the pivot 
scale $k_*$ exits the horizon for a particular choice of initial 
conditions at this point. 

Needless to say that, changing the values of the input parameters
of the model, we can easily achieve successful solutions with 
smaller values of the tensor-to-scalar ratio and that there is no 
particular fine-tuning of the parameters required in our model. 
     
\subsection{The formation of monopoles and cosmic strings}

After the termination of the first stage of inflation, the 
system moves towards the semi-shifted path and the group 
$SU(2)_{\rm R}$ breaks spontaneously to a $U(1)$ subgroup
by the nonzero value which the field $\Phi$ develops.
This leads to the formation of superheavy magnetic monopoles. 
We can obtain a rough order of magnitude estimate of the mean 
distance between the monopoles and the antimonopoles, which 
is though adequate for our purposes here. We assume that, at 
production, this distance is $p\,(2\kappa M)^{-1}$ as 
determined by the relevant Higgs boson mass with $p\sim 1$ 
being a geometric factor. In the matter 
dominated period between the two inflationary stages, this 
distance increases by a factor $\sim (\kappa^2 M^4/m^2M^2)
^{1/3}$, where $\kappa^2 M^4$ and $m^2M^2$ are the classical 
potential energy densities on the trivial and the semi-shifted 
paths, respectively. The subsequent second inflationary stage 
stretches this distance by a factor $\exp{N_2}$, where $N_2$ 
is the number of e-foldings during this stage which is large 
but not huge -- cf. Ref.~\cite{intermediate}. In the 
period of damped oscillations of the inflaton field, the 
monopole-antimonopole distance increases by another factor 
$\sim (m^2M^2/c(T_{\rm r})T_{\rm r}^4)^{1/3}$, where 
$T_{\rm r}$ is the 
reheat temperature, which we take to be $10^{9}~
{\rm GeV}$, and $c(T)=\pi^2g(T)/30$ with $g(T)$ being 
the effective number of massless degrees of freedom at cosmic 
temperature $T$. In the radiation dominated period which follows 
reheating, the monopole-antimonopole distance is multiplied 
by another factor $\sim T_{\rm r}/T\sim (4c(T)/3)^{1/4}T_{\rm r}
\sqrt{t}$, where $t$ is the cosmic time. So this distance, in 
the radiation dominated period, becomes
\beq
\sim \left(\frac{4}{3}\right)^{\frac{1}{4}} c(T_{\rm r})
^{-\frac{1}{3}}c(T)^{\frac{1}{4}}p\,
(2\kappa M)^{-1}e^{N_2}\left(\frac{\kappa^2M^4}
{T_{\rm r}^4}\right)^{\frac{1}{3}}T_{\rm r} t^{\frac{1}{2}}. 
\eeq
Equating this distance with the post-inflationary particle
horizon $\sim 2t$, we find the time $t_{\rm H}$ (in 
$t_{\rm P}\equiv m_{\rm P}^{-1}\simeq 2.7027\times 10^{-43}
~{\rm sec}$ units) at which the monopoles enter this 
horizon:
\beq
t_{\rm H}\sim\frac{p^2}{8\sqrt{3}}\,
c(T_{\rm r})^{-\frac{2}{3}}c(T_{\rm H})^{\frac{1}{2}}e^{2N_2}
\left(\frac{M}
{\kappa\, T_{\rm r}}\right)^{\frac{2}{3}},
\label{tH} 
\eeq
where $T_{\rm H}$ is the cosmic temperature at cosmic time 
$t_{\rm H}$.

The formation of monopoles is not the whole story though 
since our scenario leads to the generation of cosmic 
strings too. Indeed, after the end of the second 
inflationary stage, the system settles in one of the two 
distinct continua of SUSY vacua in Eqs.~(\ref{eq:vacua+})
and (\ref{eq:vacua-}) with $\Phi_{\pm}\simeq \pm M$ in our 
case and a linear combination of the $U(1)_{B-L}$ gauge 
symmetry and the unbroken $U(1)$ subgroup of $SU(2)_{\rm R}$ 
breaks spontaneously leading to the production of local cosmic 
strings. These strings, if they survived after recombination, 
could have a small contribution to the CMBR power spectrum 
parametrized \cite{bevis} by the dimensionless string tension 
$G\mu_{\rm s}$, where $G$ is Newton's gravitational 
constant and $\mu_{\rm s}$ is the string tension, i.e. the 
energy per unit length of the string. Applying to our case 
the results of Ref.~\cite{bevis}, which considered local 
strings within the Abelian Higgs model in the Bogomol'nyi 
limit, we write, for the string tension, 
\beq
\label{eq:stringtension}
\mu_{\rm s}=4\pi|\vev{H}|^2,
\eeq
where $\vev{H}$ is the VEV of $H$. Although the strings in our 
model are more complicated than in the Bogomol'nyi limit of the 
Abelian Higgs model, we think that the above estimate for the 
string tension is good enough for our purposes here. 

In our case, the strings decay well before recombination and, 
thus, do not affect the CMBR. The reason is that they are 
mostly open strings connecting monopoles to antimonopoles. 
This is easily understood if we realize that the breaking of 
$SU(2)_{\rm R}\times U(1)_{B-L}$ to $U(1)_Y$ by $\vev{H}$ 
and $\vev{\bar H}$ is similar to the breaking of the 
electroweak gauge group and, thus, cannot lead to any 
topologically stable monopoles or strings -- see 
Ref.~\cite{trotta} for a more detailed argument. It can 
only lead to the existence of topologically unstable dumbbell 
configurations \cite{dumbbell} consisting of an open string 
with a monopole and an antimonopole at its two ends. 

The strings, at any given time after their formation, can be 
thought of as random walks with a step of the order of the 
particle horizon \cite{walsh} -- to describe the evolution of 
this string network, we will follow closely this reference. 
They typically connect monopoles to antimonopoles, but unstable 
closed strings of limited size may also exist. As shown in 
Ref.~\cite{walsh}, at all times before the entrance of the 
monopoles into the horizon, there is of the order of one string 
segment per horizon and, thus, the ratio of the energy density 
$\rho_{\rm s}(t)$ of the string network to the total energy 
density $\rho_{\rm tot}(t)$ of the universe remains practically 
constant. At cosmic time $t_{\rm H}$, we have approximately one 
monopole-antimonopole pair per horizon volume $\sim (4\pi/3)
(2t_{\rm H})^3$ connected by an almost straight string segment 
of the size of the horizon and energy $\sim \mu_{\rm s}2
t_{\rm H}$. The energy density $\rho_{\rm s}(t_{\rm H})$ of the 
strings at cosmic time $t_{\rm H}$ is then $\sim 3G\mu_{\rm s}/
2t_{\rm H}^2$. After this time, more and more string segments 
enter the horizon, but the length of each segment remains 
constant. Consequently, the system of string segments behaves 
like pressureless matter and, thus, $\rho_{\rm s}(t)\sim 3G
\mu_{\rm s}/2(t_{\rm H}t^3)^{1/2}$, which implies that the 
`relative string energy density'
\beq
\frac{\rho_{\rm s}(t)}{\rho_\gamma(t)}\sim 2G\mu_{\rm s}
\left(\frac{t}{t_{\rm H}}\right)^{\frac{1}{2}}
\label{strenergy}
\eeq
($\rho_\gamma(t)$ is the photon energy density) 
increases with time -- remember that we are in the 
radiation dominated era of the universe and, thus, 
$\rho_{\rm tot}(t)=\rho_\gamma(t)=c(T)\,T^4=3/4t^2$. The 
strings, finally, decay at cosmic time \cite{vilenkin} 
\beq
t_{\rm d}\sim (\Gamma G\mu_{\rm s})^{-1}2t_{\rm H},
\eeq
where $\Gamma\sim 50$, by emitting gravitational waves 
with energy density $\rho_{\rm gw}(t_{\rm d})$ at 
production given by
\beq 
\label{rhogwtd}
\frac{\rho_{\rm gw}(t_{\rm d})}{\rho_\gamma
(t_{\rm d})}\sim 2\left(\frac{2}{\Gamma}\right)
^{\frac{1}{2}}(G\mu_{\rm s})^{\frac{1}{2}}
\eeq
as one can infer from Eq.~(\ref{strenergy}). Note that 
this formula also gives the maximal relative string 
energy density. 

From Eq.~(\ref{tH}) where we substitute the lowest value of 
the number of e-foldings $N_2$ and take $p=2$, we find that, 
for the two 
numerical examples presented, $t_{\rm H}\sim 4.76\times 10^{-7}~
{\rm sec}$ and $1.04\times 10^{-4}~{\rm sec}$, respectively. In 
deriving these values, we take $g(T_{\rm r})=228.75$, which 
corresponds to the spectrum of the minimal SUSY SM, and  
$g(T_{\rm H})=40.75$ and $10.75$ in our two numerical examples, 
respectively. The above values of $g(T_{\rm H})$ are consistent 
with the effective number of massless degrees of freedom at the 
cosmic temperatures $T_{\rm H}$ corresponding to the values of 
the cosmic time $t_{\rm H}$ obtained. We see that the strings 
enter the horizon well before the time of big bang 
nucleosynthesis. Their decay time is $t_{\rm d}\sim 5.97\times 
10^{-2}~{\rm sec}$ and $5.49~{\rm sec}$, in the two cases, as 
one can find from the corresponding dimensionless string 
tensions 
\beq
\label{Gmus}
G\mu_{\rm s}=\frac{|\vev{H}|^2}{2}\simeq \frac{mM}{2\lambda} 
\simeq 3.19 \times 10^{-7}\quad{\rm and}\quad 7.57\times 10^{-7}.
\eeq
This means that the cosmic strings decay around the time of 
nucleosynthesis and, thus, well before recombination which 
takes place at a cosmic time $\sim 10^{13}~{\rm sec}$. As a 
consequence, they do not affect the CMBR. Their maximal relative 
energy density in the universe is $\sim 2.26\times 10^{-4}$ and 
$3.48\times 10^{-4}$ for our two numerical examples. So the 
cosmic strings remain always subdominant. In particular, they do 
not disturb nucleosynthesis at all. 

Had the strings been around until the present time, an upper 
bound would have to be imposed on the dimensionless string 
tension in order 
to keep their contribution to the CMBR power spectrum at an 
acceptable level. For the Abelian-Higgs field theory model, 
this bound is found to be \cite{strings}
\beq
\label{Gmu}
G\mu_{\rm s}\lesssim 3.2\times 10^{-7}.
\eeq  
In our first numerical example, the dimensionless string 
tension $G\mu_{\rm s}$ given in 
Eq.~(\ref{Gmus}) almost saturates the upper bound in 
Eq.~(\ref{Gmu}), but violates the recent more stringent 
upper bound \cite{nano} 
\beq
\label{Gmunano}
G\mu_{\rm s}\lesssim 3.3\times 10^{-8}
\eeq
from pulsar timing arrays, which also holds for strings 
surviving until the present time. Our second numerical 
example violates both the bounds in Eqs.~(\ref{Gmu}) and 
(\ref{Gmunano}). Thus, both our examples are only possible 
because the strings decay sufficiently early.

The value of the ratio of the energy density of the gravitational 
waves produced by the strings to that of the photons at the 
present cosmic time $t_0$ is found from Eq.~(\ref{rhogwtd}) 
to be -- cf. Ref.~\cite{maggiore} --
\beq 
\frac{\rho_{\rm gw}(t_0)}{\rho_\gamma(t_0)}\sim 2
\left(\frac{2}{\Gamma}\right)^{\frac{1}{2}}
(G\mu_{\rm s})^{\frac{1}{2}}\left(\frac{3.9}{10.75}
\right)^\frac{4}{3}.
\eeq
The present abundance of these gravitational waves is then 
given by
\beq
\Omega_{\rm gw} h^2 (t_0)\sim \left(\frac{\rho_{\rm gw}(t_0)}
{\rho_\gamma(t_0)}\right)\left(\frac{\rho_\gamma(t_0)}
{\rho_{\rm c}(t_0)}\right)h_0^2\, ,
\eeq
where $\rho_{\rm c}(t_0)$ is the present critical energy density 
of the universe and $h_0\simeq 0.7$ the present value of the Hubble 
parameter in units of ${\rm km~sec^{-1}~Mpc^{-1}}$. We find that, 
for our two numerical examples, $\Omega_{\rm gw} h^2 (t_0)\sim 
2.18\times 10^{-9}$ and $3.35\times 10^{-9}$, respectively. The 
frequency $f(t_{\rm d})$ of these gravitational waves at production 
must be $\sim t_{\rm H}^{-1}$ since the length of the decaying 
strings is $\sim 2t_{\rm H}$ -- see Ref.~\cite{vilenkin}. The 
present value of this frequency is then 
\beq
f(t_0)\sim t_{\rm H}^{-1}\left(\frac{t_d}{t_{\rm eq}}\right)^
{\frac{1}{2}}\left(\frac{t_{\rm eq}}{t_0}\right)^{\frac{2}{3}},
\eeq  
where $t_{\rm eq}$ is the equidensity time at which matter starts
dominating the universe. For the two numerical examples, this 
frequency turns out to be $\sim 1.06\times 10^{-4}~{\rm Hz}$ and 
$4.68\times 10^{-6}~{\rm Hz}$, respectively. Note that the 
estimate of the frequencies and abundances of the gravity waves 
presented here cannot be made much more accurate, but it can 
certainly be considered good enough for our purposes. 

We see that the predicted frequencies of the gravitational 
waves are too high to yield any restrictions from CMBR 
considerations \cite{maggiore}. They are also well above 
the range of frequencies probed by the pulsar timing array 
observations \cite{liu}. So, the most recent stringent 
bound \cite{nano} from these observations does not apply 
to our case. However, the frequency of the gravitational 
waves in our first numerical example lies marginally within 
the range to be probed by the future space-based laser 
interferometer gravitational-wave observatories such as 
the evolved laser interferometer space antenna/new 
gravitational-wave observatory (eLISA/NGO) \cite{lisa}, 
which is expected to be able to detect values of 
$\Omega_{\rm gw} h^2 (t_0)$ as low as $4\times 10^{-10}$.  
Our overall conclusion is that the monopole-string system 
disappears without causing any trouble, but the gravitational 
waves that it generates may be probed by future space-based 
laser interferometer observations.  


\section{Conclusions}
\label{sec:concl}

In view of the recent results \cite{joint1,joint2} indicating 
that appreciable values of the tensor-to-scalar ratio in the 
CMBR cannot be excluded, we addressed the question 
whether such values can be obtained in SUSY hybrid inflation 
models resulting from particle physics. To this end, we have 
considered a reduced version of the extended SUSY PS model of 
Ref.~\cite{quasi}, which was initially constructed for solving 
the $b$-quark mass problem of the simplest SUSY PS model with 
universal boundary conditions. The reason for focusing on this 
model is that it is known to support successful versions of 
hybrid inflation like the standard-smooth one \cite{stsmhi}. 
This scenario is compatible with all the recent data even with 
a minimal K\"{a}hler potential, but predicts negligible values 
of the tensor-to-scalar ratio. 

In the context of this particular particle physics model, we 
demonstrated that a two stage hybrid inflationary scenario 
which can predict values of the tensor-to-scalar ratio of the 
order of ${\rm few}\times 10^{-2}$ can be constructed. For 
the values of the parameters considered in this paper, 
the model in the global SUSY limit possesses practically two 
classically flat directions, the trivial and the 
semi-shifted \cite{semi} one. The SUGRA corrections to the 
potential stabilize the trivial flat direction so that it 
becomes able to support a first stage of inflation. All the 
cosmological scales exit the horizon during this inflationary 
stage and our present horizon undergoes a limited number of 
e-foldings. The tensor-to-scalar ratio can acquire appreciable 
values as a result of sufficiently mild (i.e. non-catastrophic), 
but still appreciable 
SUGRA corrections combined with strong radiative corrections 
to the inflationary potential, while the value of the scalar 
spectral index remains acceptable. We can obtain values of the 
tensor-to-scalar ratio up to about $0.05$. Larger values would 
require unacceptably large running of the scalar spectral index. 
 
The additional number of e-foldings required for solving 
the standard problems of hot big bang cosmology are 
generated by a second inflationary stage taking place 
along the semi-shifted path, where $U(1)_{B-L}$ is 
unbroken. This is possible since the 
semi-shifted direction, being almost orthogonal to the 
trivial path, is not affected by the strong radiative 
corrections on the trivial path and also because the SUGRA
corrections on the semi-shifted path remain mild.

At the end of the first inflationary stage, the group 
$SU(2)_{\rm R}$ breaks spontaneously to a $U(1)$ subgroup 
and, thus, magnetic monopoles are formed. The subsequent 
spontaneous 
breaking of a linear combination of this $U(1)$ and 
$U(1)_{B-L}$ at the end of the second inflationary 
stage leads to the production of cosmic strings 
connecting these monopoles to antimonopoles. At later 
times, the monopoles enter the horizon and the 
string-monopole system decays into gravity waves well 
before recombination without leaving any trace in the 
CMBR. The resulting gravity waves, however, may be 
measurable in the future.   

\def\ijmp#1#2#3{{Int. Jour. Mod. Phys.}
{\bf #1},~#3~(#2)}
\def\plb#1#2#3{{Phys. Lett. B }{\bf #1},~#3~(#2)}
\def\zpc#1#2#3{{Z. Phys. C }{\bf #1},~#3~(#2)}
\def\prl#1#2#3{{Phys. Rev. Lett.}
{\bf #1},~#3~(#2)}
\def\rmp#1#2#3{{Rev. Mod. Phys.}
{\bf #1},~#3~(#2)}
\def\prep#1#2#3{{Phys. Rep. }{\bf #1},~#3~(#2)}
\def\prd#1#2#3{{Phys. Rev. D }{\bf #1},~#3~(#2)}
\def\npb#1#2#3{{Nucl. Phys. }{\bf B#1},~#3~(#2)}
\def\np#1#2#3{{Nucl. Phys. B }{\bf #1},~#3~(#2)}
\def\npps#1#2#3{{Nucl. Phys. B (Proc. Sup.)}
{\bf #1},~#3~(#2)}
\def\mpl#1#2#3{{Mod. Phys. Lett.}
{\bf #1},~#3~(#2)}
\def\arnps#1#2#3{{Annu. Rev. Nucl. Part. Sci.}
{\bf #1},~#3~(#2)}
\def\sjnp#1#2#3{{Sov. J. Nucl. Phys.}
{\bf #1},~#3~(#2)}
\def\jetp#1#2#3{{JETP Lett. }{\bf #1},~#3~(#2)}
\def\app#1#2#3{{Acta Phys. Polon.}
{\bf #1},~#3~(#2)}
\def\rnc#1#2#3{{Riv. Nuovo Cim.}
{\bf #1},~#3~(#2)}
\def\ap#1#2#3{{Ann. Phys. }{\bf #1},~#3~(#2)}
\def\ptp#1#2#3{{Prog. Theor. Phys.}
{\bf #1},~#3~(#2)}
\def\apjl#1#2#3{{Astrophys. J. Lett.}
{\bf #1},~#3~(#2)}
\def\apjs#1#2#3{{Astrophys. J. Suppl.}
{\bf #1},~#3~(#2)}
\def\n#1#2#3{{Nature }{\bf #1},~#3~(#2)}
\def\apj#1#2#3{{Astrophys. J.}
{\bf #1},~#3~(#2)}
\def\anj#1#2#3{{Astron. J. }{\bf #1},~#3~(#2)}
\def\mnras#1#2#3{{MNRAS }{\bf #1},~#3~(#2)}
\def\grg#1#2#3{{Gen. Rel. Grav.}
{\bf #1},~#3~(#2)}
\def\s#1#2#3{{Science }{\bf #1},~#3~(#2)}
\def\baas#1#2#3{{Bull. Am. Astron. Soc.}
{\bf #1},~#3~(#2)}
\def\ibid#1#2#3{{\it ibid. }{\bf #1},~#3~(#2)}
\def\cpc#1#2#3{{Comput. Phys. Commun.}
{\bf #1},~#3~(#2)}
\def\astp#1#2#3{{Astropart. Phys.}
{\bf #1},~#3~(#2)}
\def\epjc#1#2#3{{Eur. Phys. J. C}
{\bf #1},~#3~(#2)}
\def\nima#1#2#3{{Nucl. Instrum. Meth. A}
{\bf #1},~#3~(#2)}
\def\jhep#1#2#3{{J. High Energy Phys.}
{\bf #1},~#3~(#2)}
\def\jcap#1#2#3{{J. Cosmol. Astropart. Phys.}
{\bf #1},~#3~(#2)}
\def\lnp#1#2#3{{Lect. Notes Phys.}
{\bf #1},~#3~(#2)}
\def\jpcs#1#2#3{{J. Phys. Conf. Ser.}
{\bf #1},~#3~(#2)}
\def\aap#1#2#3{{Astron. Astrophys.}
{\bf #1},~#3~(#2)}
\def\mpla#1#2#3{{Mod. Phys. Lett. A}
{\bf #1},~#3~(#2)}

\end{document}